\documentclass[amsmath,preprint,prb,11pt,nobibnotes]{revtex4-1}

\usepackage{graphicx}
\usepackage{bm}
\usepackage[mathscr]{euscript}

\usepackage{hyperref}
\hypersetup{breaklinks=true}
\usepackage{subfigure}

\begin{document}

\title{A hybrid stochastic hierarchy equations of motion approach to treat the
   low temperature dynamics of non-Markovian open quantum systems}

\author{
       Jeremy M. Moix
       } 
\author{
       Jianshu Cao
       } \email{jianshu@mit.edu}
\affiliation{
       Department of Chemistry,
       Massachusetts Institute of Technology,
       77 Massachusetts Avenue, Cambridge, MA 02139
      }

\date{\today}

\begin{abstract}

The hierarchical equations of motion technique has found widespread success
as a tool to generate the numerically exact dynamics of 
non-Markovian open quantum systems.
However, its application to low temperature environments remains
a serious challenge due to the need for a deep hierarchy that arises from 
the Matsubara expansion of the bath correlation function.
Here we present a hybrid stochastic hierarchical equation of motion 
(sHEOM) approach that alleviates this bottleneck and leads to a numerical cost
that is nearly independent of temperature.
Additionally, the sHEOM method generally converges 
with fewer hierarchy tiers allowing for the treatment of larger systems.
Benchmark calculations are presented on the dynamics of two level systems 
at both high and low temperatures to demonstrate the efficacy of the approach.
Then the hybrid method is used to generate the exact dynamics
of systems that are nearly impossible to treat by the standard hierarchy.
First, exact energy transfer rates are calculated across a broad range
of temperatures revealing the deviations from the F{\"o}rster rates.
This is followed by computations of the entanglement dynamics
in a system of two qubits at low temperature spanning the
weak to strong system-bath coupling regimes.

\end{abstract}

\maketitle

\section{Introduction}

Despite a long history and intense interest, there are still relatively
few methods capable of generating the numerically exact 
dynamics of open quantum systems across a wide range of system parameters.
Among the most successful numerical methods are those derived from the path
integral formalism, such as the hierarchical equations of 
motion (HEOM),\cite{Tanimura2006,Ishizaki2005} 
the quasi-adiabatic path integral (QUAPI) approach,\cite{Makri1995a,makri98} 
or direct integration of the path integral expression through Monte Carlo 
methods.\cite{egger00,muhlbacher03}
The widespread success of the path integral formalism in open quantum systems
lies in the ability to analytically integrate out the 
(Gaussian) environmental degrees of freedom.
This procedure leads to the Feynman-Vernon influence functional
which accounts for all of the effects of the bath 
on the system.\cite{feynman63,grabert88}
The price to be paid for this simplification, however, is that the 
influence functional is non-local in time, with a
correlation time that depends on the intrinsic relaxation time of the bath,
the temperature, and the system-bath coupling strength.
The key insight in developing the QUAPI formalism was that,
in many cases, the bath correlation function, and hence the influence
functional itself, decays relatively quickly so that the propagation may be
obtained through a small number of deterministic tensor multiplications.
Alternatively the HEOM replaces the influence functional by a 
set of auxiliary density matrices that account for the non-Markovian effects
of the bath, provided that the spectral density may be represented in 
Drude-Lorentz form. 
Essentially, both the HEOM and QUAPI represent expansions of the influence 
functional in terms of the memory time of the environment, 
and as a result they often become prohibitively expensive for 
strong system-bath interactions, highly non-Markovian environments, 
or low temperatures.

The main result of this work is a numerical algorithm that significantly
extends the parameter regimes accessible to the HEOM, although the general
approach outlined here is equally applicable to QUAPI.
As is well known, the quantum bath correlation function that appears in the
influence functional is a complex function with real and imaginary parts.
The real part is responsible for the fluctuations from the environment
with a magnitude that increases with temperature,
whereas the imaginary part is temperature independent and accounts for 
the dissipative effects of the environment.
All of the methods mentioned above treat these two components on the
same footing, but this restriction is not necessary or indeed desirable.
Here we show that it can be advantageous to treat the real 
and imaginary parts of the correlation function by different methods.
The central idea is to perform a stochastic unraveling of the real part
of the bath correlation function appearing in the influence functional.
The remaining imaginary term will then be treated by the HEOM.

The stochastic unraveling of the influence functional has recently 
formed the basis of a new approach for computing the exact dynamics of open
quantum systems.\cite{stockburger02,shao04,cao96,moix12}
The unraveling is achieved through a series of Hubbard-Stratonovich
transformations that ultimately result in two coupled linear stochastic 
differential equations describing the respective forward- and 
backward-time propagators.
As a Monte Carlo, wavefunction-based approach, this scheme appears quite 
promising for computing the exact dynamics of large systems.
Performing a Hubbard-Stratonovich transformation only on the real part 
of the bath correlation function in the influence functional 
generates an auxiliary field that is purely real.
If one stops at this point and ignores the remaining imaginary part of 
the bath correlation function, then this procedure
leads to an evolution equation identical to that of the Haken-Strobl model,
except that the Gaussian noise is colored instead of white.
In fact, numerical simulations in this case are no more difficult 
than the Haken-Strobl model itself, and can be readily applied to very 
large systems.\cite{zhong11,chen13}
However, due to the neglect of the dissipative aspects of the environment,
detailed balance is recovered only at infinite temperature.

Correctly treating the imaginary part of the correlation function appearing 
in the influence functional by a stochastic unraveling is more complicated.
It requires a complex auxiliary field that introduces a corresponding
complex noise term into the stochastic evolution equations.\cite{strunz96}
Consequently the norm of the individual realizations of the wavefunction
is not conserved during the propagation.
Of course the ensemble average is perfectly normalized, but this
norm loss at the realization level greatly degrades the convergence 
properties of the Monte Carlo procedure.
In principle, these difficulties may be circumvented through the use 
of a Girsanov transformation and other numerical schemes, 
but the resulting stochastic Schr{\"o}dinger equations are generally 
nonlinear.\cite{stockburger02,shao04}

Here, we present an approach that essentially combines the 
strengths of the stochastic and deterministic methods,
while mitigating their respective difficulties.
The central idea is to perform a stochastic unraveling only of the 
real part of the bath correlation function appearing in the influence 
functional.
The remaining imaginary term can then be treated by QUAPI or the HEOM.
In fact, similar ideas have been previously proposed, 
although generally coupled with further approximations. 
For example, Stockburger and Mak have applied this approach
in the context of QUAPI, although the numerical implementation 
was restricted to an environment that was nearly 
Markovian.\cite{stockburger98}
In subsequent work, they treated the real part of the 
influence functional exactly through a Hubbard-Stratonovich transformation, 
as is also used here, but a Markovian approximation was made for the 
dissipative term.\cite{stockburger99}
Although approximate, this approach leads to a single, closed equation of 
motion for the reduced density matrix that resembles Kubo's stochastic 
Liouville equation. 
Additionally, Tanimura outlined such an approach within 
the HEOM formalism in Ref.~\onlinecite{Tanimura2006} which was referred to as
the Fokker-Planck equation with Langevin forces, 
but the proposed algorithm and preliminary numerical 
results were only valid in the high temperature limit. 
A similar result was also derived from a purely stochastic perspective in 
Ref.~\onlinecite{zhou05}.
While the formal results in Refs.~\onlinecite{Tanimura2006,zhou05}
are exact in principle, the simple description of the 
real and imaginary parts of the correlation function by
stochastic and deterministic schemes, respectively, 
leads to an unstable numerical algorithm which has limited their practical
application.
Through the introduction of a reference temperature, here,
we propose an improved decomposition scheme 
that leads to a substantial improvement over the previous methods.
It is shown that a suitable choice of the reference temperature can provide 
an optimal balance between the stochastic sampling and deterministic evolution.
We make no approximations and treat the non-Markovian 
characteristics of the bath exactly, demonstrating that the hybrid
stochastic hierarchy approach is widely applicable across a broad range 
of system parameters including, in particular, 
the low temperature, strong coupling regime.

As with the standard HEOM, the spectral density of the bath is restricted 
to be of Drude-Lorentz form. 
However, this restriction offers many advantages.
In this case, the imaginary part of the bath correlation function 
consists of a single, temperature-independent exponential term.
Thus, the only convergence parameter with respect to the depth of hierarchy 
is determined by the reorganization energy and the time scale of the
bath --the infinite summation over the Matsubara frequencies is 
performed exactly through a Monte Carlo sampling of the auxiliary stochastic 
field.
The real power of this approach over the standard HEOM is that it 
is valid for arbitrary temperatures with 
a numerical cost that is nearly independent of temperature.
In addition, because the imaginary part of the bath correlation function
is generally of smaller magnitude than the real part and also decays more 
rapidly,
the stochastic hierarchy can typically be truncated at a much lower tier
than with the standard approach.
Thus, in principle, one can treat larger systems with the hybrid approach.
The only drawback is that a stochastic average over the 
independent HEOM evolutions is required.
However, our calculations so far have indicated that the stochastic 
average generally converges rapidly, and of course, Monte Carlo algorithms 
can be trivially parallelized.

In the following section, the hybrid stochastic hierarchy equations
of motion (sHEOM) are formalized.
Following this, benchmark calculations of two-level systems 
are presented in Sec.~\ref{sec:results} 
for which standard hierarchy results can be obtained. 
Then energy transfer rates in a model donor-acceptor system 
are computed across a broad range of temperatures.
Finally, the entanglement dynamics in a system 
of two qubits are presented at near zero temperature 
spanning the weak to strong system-bath coupling regimes.

\section{Formalism}
\subsection{Preliminaries}

For clarity we consider only one-dimensional systems with continuous
degrees of freedom.
For the case of discrete systems, the path integral formalism is most 
readily developed either through the use of Grassmann 
variables\cite{Ishizaki2005} or in the mapping representation.\cite{novikov04}
However, our final working expressions are valid for either continuous or 
discrete systems and numerical results will be presented for the latter.
We consider generic system-bath Hamiltonians of the form, 
$\hat H = \hat H_0 + \hat H_{sb}$ where
\begin{align}
   \hat H_{sb} & = \frac12\sum_j \left[\hat p_j^2 + \omega_j^2 
      \left(\hat x_j^2 - \frac{c_j}{\omega_j^2} \hat V(\hat Q)\right)^2 \right]
      \;.
\end{align}
The specific form of the system Hamiltonian, $\hat H_0$, is irrelevant at 
this point and need not be specified.
The thermal bath is composed of independent harmonic oscillators 
characterized by their respective mass-weighted coordinates, $\hat x_j$,
and momenta, $\hat p_j$,
with frequencies $\omega_j$ and coupling constants $c_j$ to the system
degrees of freedom through the interaction potential $\hat V(\hat Q)$,
where $\hat Q$ denotes a system coordinate.

Assuming that the bath remains in thermal equilibrium
and that the initial state of the composite system factorizes such that
$\rho(0) = \rho_s(0) \rho_b(0)$,
then the environmental degrees of freedom may be integrated out 
analytically.\cite{feynman63,grabert88}
As a result the path integral expression for the 
forward-backward propagation of the reduced density matrix 
is succinctly given by
\begin{equation}
   U(x_f, y_f,t; x_0, y_0,0) = 
   \int_{x(0)=x_0}^{x(t)=x_f}  {\mathscr D}[x]
   \int_{y(0)=y_0}^{y(t)=y_f}  {\mathscr D}[y]
   \exp\left(\frac{i}{\hbar}\left(S_0[x] - S_0[y]\right)\right)
   F[x,y] \;,
   \label{eq:feynmann}
\end{equation}
where $S_0[x]$ denotes the classical action function associated with 
the bare system Hamiltonian, $H_0$, computed along the path starting 
at $x_0$ at $t=0$ and ending at $x_f$ at time $t$.
All of the effects of the environment on the system dynamics 
are accounted for by 
the Feynman-Vernon influence functional,\cite{feynman63} which,
for future convenience, is decomposed as $F[x,y] = F_r[x,y]F_i[x,y]F_b[x,y]$ 
where the respective terms are associated with the real part of the bath 
correlation function ($F_r[x,y]$), the imaginary part ($F_i[x,y]$), and the
static bath renormalization ($F_b[x,y]$).
Each term is given explicitly as
\begin{align}
   F_r[x,y] & = \exp\left(-\frac{1}{\hbar} \int_0^t dt' \int_0^{t'} dt''
   V^\times(t') K_r(t'-t'')V^\times(t'') \right) 
    \;, \nonumber \\
   F_i[x,y] & = \exp\left(-\frac{i}{\hbar} \int_0^t dt' \int_0^{t'} dt''
    V^\times(t') K_i(t'-t'')V^\circ(t'')\right) 
     \;, \nonumber \\
   F_b[x,y] & = \exp\left(-\frac{i}{\hbar} \int_0^t dt' \;
                   \lambda V^\times(t')V^\circ(t')\right) \;,
   \label{eq:influence_functional}
\end{align}
where the notation $V^\times(t) = V(x(t))-V(y(t))$ and 
$V^\circ(t)=V(x(t))+V(y(t))$ has been introduced, along with
the decomposition of the bath correlation function into its
real, $K_r$, and imaginary, $K_i$, parts,
\begin{align}
   K(t) & = K_r(t) + i K_i(t)\\
        & = \int_0^\infty \frac{d\omega}{\pi}\: J(\omega) 
        \left[ \coth\left(\frac{\hbar\beta\omega}{2}\right)\cos(\omega t) -i \sin(\omega t) \right]
   \;. \label{eq:correlation}
\end{align}
The spectral density function,
\begin{equation}
   J(\omega) = \frac{\pi}{2} \sum_{j=1} \frac{c_j^2}{\omega_j} 
               \delta\left(\omega-\omega_j\right)\;.
\end{equation}
contains all of the relevant features in terms of the system dynamics.

The HEOM requires that the bath is described by the Drude-Lorentz 
spectral density,
\begin{equation}
   J_D(\omega) = 2\lambda\omega_c \frac{\omega}{\omega^2 + \omega_c^2}
   \;.
\end{equation}
where $\omega_c$ is the cutoff frequency of the bath and 
the reorganization energy $\lambda$ is defined as
\begin{equation}
   \lambda = \frac12 \sum_{j=1} \frac{c_j^2}{\omega_j^2} 
   = \frac{1}{\pi} \int_0^\infty d\omega \; \frac{J(\omega)}{\omega}
   \;,
\end{equation}
where the second equality is obtained in the continuum limit.
For the Drude spectrum, the corresponding bath correlation function may be
computed analytically and the time dependence is of multi-exponential form,
\begin{align}
   K_r(t) & = \lambda\omega_c \cot\left(\hbar \beta\omega_c/2\right) 
              e^{-\omega_c t}
            + \frac{4\lambda\omega_c}{\hbar \beta}
              \sum_{j=1}^\infty \frac{\nu_j}{\nu_j^2 - \omega_c^2} e^{-\nu_j t} 
              \nonumber \\
   K_i(t) & = -\lambda\omega_c e^{-\omega_c t} 
   \;, \label{eq:drude}
\end{align}
where the Matsubara frequencies are defined as 
$\nu_j = \frac{2\pi j}{\hbar \beta}$,

With these preliminaries, the formulation of the sHEOM is 
rather straightforward.
In the next subsection, a Hubbard-Stratonovich transformation 
is first performed on $F_r[x,y]$, the influence functional containing the real
part of the bath correlation function.\cite{stockburger99,schul86,moix12}
Then in Sec.~\ref{sec:heom} the remaining imaginary part is developed 
in a hierarchical expansion.
Ultimately the expressions for the time evolution of the reduced density
matrix are formally identical to those of the standard HEOM, except 
that here the Hamiltonian is stochastic.\cite{Tanimura2006,zhou05}

\subsection{Partial Stochastic Unraveling}

The Hubbard-Stratonovich transformation is a Gaussian integral identity.
Physically, it allows for the decomposition of an interacting system into
separate, non-interacting systems but subject to the influence of a common 
auxiliary field.\cite{hubbard59}
We have recently employed such an approach in the case of the 
equilibrium reduced density matrix which led to a highly
efficient numerical algorithm.\cite{lee12,lee12a,moix12}
For real time dynamics, this transformation was applied to
the entire influence functional in Ref.~\onlinecite{stockburger02} leading
to a set of stochastic Schr{\"o}dinger equations.
Here we apply the transformation only to the real part of the correlation
function in the influence functional as in 
Ref.~\onlinecite{stockburger99,Tanimura2006}, 
which introduces the corresponding real, auxiliary field, $\xi(t)$. 
As such, the term $F_r[x,y]$ is equivalently represented as
\begin{align}
   F_r[x,y]
   & = \int {\mathscr D}[\xi]\:
   N_{\xi} \exp\left(-\frac{1}{2\hbar} \int_0^t dt' \int_0^t dt'' \: 
         \xi(t') K_r^{-1}(t'-t'') \xi(t'') 
        + \frac{i}{\hbar}\int_0^t dt'\xi(t')V^\times(t')  \right) \;,
        \nonumber \\ 
   & = \int {\mathscr D}[\xi]\:
   P[\xi]\exp\left(\frac{i}{\hbar}\int_0^t dt'\xi(t')V^\times(t')  \right) \;,
        \nonumber \\ 
   & = \left \langle 
   \exp\left(\frac{i}{\hbar}\int_0^t dt'\xi(t')V^\times(t')  \right) 
   \right\rangle_{\xi}
   \;,  \label{eq:symmetric_decomp}
\end{align}
where $N_\xi$ denotes the normalization of the Gaussian functional integral.
The Hubbard-Stratonovich transformation is valid provided that $K_r(t)$ is
positive semi-definite, which is ensured
in this case since it is defined by a covariance function of the bath.
Performing the Gaussian integration over $\xi(t)$ clearly leads back to the
original expression for $F_r[x,y]$ in Eq.~\ref{eq:influence_functional}.
The normalization of the functional integral implies that 
the Gaussian term in Eq.~\ref{eq:symmetric_decomp} 
can serve as a true probability measure, $P[\xi]$,
which leads to the subsequent equalities in Eq.~\ref{eq:symmetric_decomp}.
These features allow for the interpretation of $\xi(t)$ as 
a colored noise driving the dynamics
with one- and two-time correlation functions that obey the relations,
\begin{align}
   \left \langle \xi(t) \right \rangle &=  0 \nonumber \\ 
   \left \langle \xi(t)\xi(t') \right \rangle &= \hbar K_r(t-t')
   \label{eq:noise_statistics}\;.
\end{align}

Since the unraveled influence functional in Eq.~\ref{eq:symmetric_decomp} 
and the bath renormalization in Eq.~\ref{eq:influence_functional} are 
both local in time, 
they can be written as additional action terms such that the partially 
unraveled propagator is given by,
\begin{multline}
   U(x_f, y_f,t; x_0, y_0,0)   = \\ 
     \left \langle 
    \int_{x(0)=x_0}^{x(t)=x_f}  {\mathscr D}[x]
    \int_{y(0)=y_0}^{y(t)=y_f}  {\mathscr D}[y]
    \exp\left(\frac{i}{\hbar}\left(S_0[x] - S_0[y]\right)\right)
    \exp\left(\frac{i}{\hbar}\left(S_1[x] - S_1[y]\right)\right)
    F_i[x,y] \right \rangle_{\xi} \;,
\end{multline}
where $S_0$ is the classical action function associated with the bare
system Hamiltonian $H_0$. 
The noise and potential renormalization are included in the additional action
term,
\begin{equation}
   S_1[x] = \int_0^t dt'\: \xi(t') V(x(t')) - \lambda V(x(t'))^2 \;.
\end{equation}
Alternatively, it is also apparent that a modified action 
$\tilde S[x] = S_0[x] + S_1[x]$ can be defined that is 
generated from the corresponding time-dependent (stochastic) Hamiltonian, 
\begin{equation}
   \tilde H(\xi;t) = \hat H_0 - \xi(t) \hat V(\hat x) + \lambda \hat V(\hat x)^2
   \;.
   \label{eq:stochastic_H}
\end{equation}
This interpretation will be used in the ensuing developments.

In a similar manner one may perform an additional Hubbard-Stratonovich 
transformation on the remaining portion of the influence functional, 
$F_i[x,y]$, that ultimately leads to the stochastic Schr{\"o}dinger equations
of Ref.~\onlinecite{stockburger02}.
However, in this case the auxiliary stochastic fields are required to be 
complex which leads to serious numerical difficulties due to the 
associated loss of norm of the individual wavefunction 
realizations.\cite{strunz96}
Therefore, we stop here while the auxiliary field is well behaved
and develop $F_i[x,y]$ in a hierarchical expansion following
Ref.~\onlinecite{Tanimura2006}.
It should be mentioned, however, that the QUAPI formalism could be equally
applied to the imaginary part of the bath correlation as proposed in 
Ref.~\onlinecite{stockburger98}.

\subsection{Hierarchical expansion} \label{sec:heom}

Specifying to the Drude-Lorentz spectral density, then the 
time dependence of the bath correlation function is of a 
purely exponential form (see Eq.~\ref{eq:drude}).
For such correlation functions, the hierarchical equations of motion
have proved to be a highly efficient numerical approach
to compute the exact dynamics of open quantum systems at moderately 
high temperature.
The standard approach, however, becomes extremely costly at low temperature.
The source of this difficulty is that the temperature dependence of 
the real part of the bath correlation is expressed as an infinite
summation over Matsubara terms.
At low temperatures, many terms of the series must be retained 
which gives rise to a very deep hierarchy of equations before truncation
is acceptable,
although some  progress has been made recently by utilizing a Pad{\'e} 
decomposition of the bath correlation function.\cite{ding11}
The stochastic approach developed in the previous subsection  
circumvents this difficulty entirely by replacing the temperature-dependent 
terms in the influence functional by a noise sampling procedure that is 
independent of the number of Matsubara frequencies. 
The remaining imaginary part of the correlation function appearing 
in $F_i[x,y]$ consists of only a single exponential term
and is independent of temperature.

At this point, one may follow the standard derivations of 
the hierarchy expansion, leading to nearly identical results
for the equations of motion of the auxiliary density 
matrices.\cite{Tanimura2006,Ishizaki2005}
For conciseness, this procedure will not be 
reproduced here and we only present the final result.
For a given realization of the stochastic noise, $\xi(t)$, 
the hierarchy equations of motion are given by,
\begin{align}
   \frac{\partial}{\partial t} \hat \rho_n(\xi; t) 
      & = -\left(\frac{i}{\hbar} \tilde H(\xi; t)^\times + n \omega_c\right) \hat \rho_n(\xi; t)
      -\frac{i}{\hbar}\hat V^\times \hat \rho_{n+1}(\xi; t)
          - n \frac{\lambda \omega_c}{\hbar}
                  \hat V^\circ \hat \rho_{n-1}(\xi; t)
      \label{eq:hierarchy}
\end{align}
where  the hyper-operator notations 
$\hat V^\times \hat O = \hat V\hat O - \hat O\hat V$ and 
$\hat V^\circ \hat O = \hat V\hat O + \hat O \hat V$ denote 
commutators and anti-commutators, respectively. 
The stochastic Hamiltonian is given in Eq.~\ref{eq:stochastic_H},
and the associated noise statistics are given in Eq.~\ref{eq:noise_statistics}.
The numerically exact reduced density matrix is obtained after preforming 
the stochastic average, 
$\rho(t) = \left \langle \rho_0(\xi;t) \right \rangle_\xi$.
The generalization to systems containing multiple independent baths
is identical to those in Refs.~\onlinecite{Tanimura2006,Ishizaki2005}
except that the stochastic Hamiltonians must also include multiple 
independent noise terms.
Eq.~\ref{eq:hierarchy} has been previously derived by
Tanimura in Ref.~\onlinecite{Tanimura2006} (cf.~Eq~6.35 therein),
and referred to as the Fokker-Planck equation with Langevin forces.
Similarly it was also derived from a purely stochastic approach in 
Ref.~\onlinecite{zhou05} and used to study
the low temperature dynamics of the spin-boson model.
The closed stochastic equation for the density matrix obtained
by Stockburger and Mak is recovered when a Markovian approximation is made 
to the imaginary part of the influence functional,
in which case the hierarchy truncates at the lowest level.\cite{stockburger99}

\subsection{Reference Temperature}

\begin{figure}
   \includegraphics*[width=0.7\textwidth]{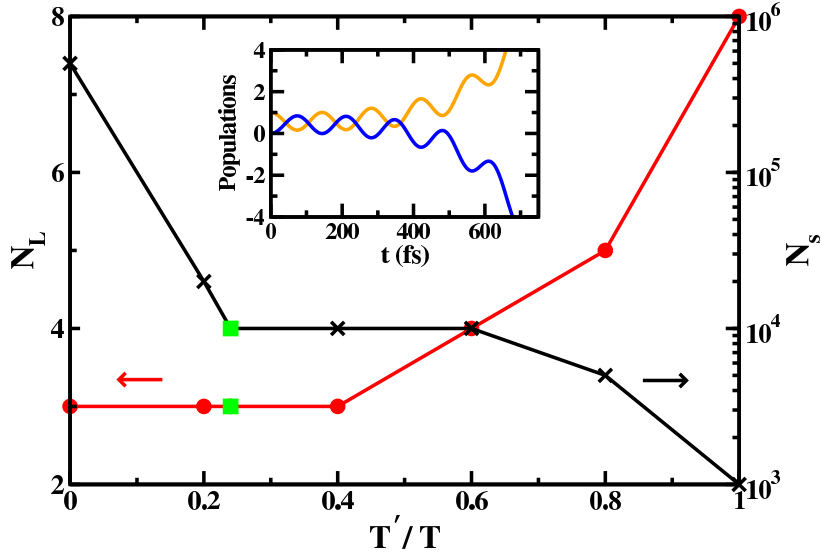}
   \caption{
      The convergence properties of the sHEOM 
      as a function of the reference temperature $T'$
      in an unbiased two level system coupled to two independent, identical
      baths at $T=300$ K with electronic coupling,
      reorganization energy, and cutoff frequency of $J=100$ cm$^{-1}$,
      $\lambda= 2J$, and $\omega_c=J$, respectively.
      The axis labels $N_L$ and $N_s$ refer to the number of hierarchy
      levels and Monte Carlo samples required for convergence, respectively.
      The green square symbols denote the choice $\beta' = 2/\omega_c$
      used here.
      The inset displays the loss of positivity in the noiseless
      dynamics of Eq.~\ref{eq:hierarchy} ($T'/T=0)$, which causes the
      sharp increase in the required number of Monte Carlo samples.
   }
   \label{fig:scaling}
\end{figure}
Unfortunately there is a formidable numerical instability associated 
with the hierarchy as given in Eq.~\ref{eq:hierarchy} that has limited
its practical application.
If one considers the noiseless case with $\xi(t)=0$, then the resulting
deterministic hierarchy equations of motion describe a purely dissipative
quantum bath.
There is no physical limit in which such a case occurs since
thermal fluctuations are always present, even at zero temperature.
As a result the reduced density matrix obtained from the noiseless 
version of Eq.~\ref{eq:hierarchy} is not necessarily positive semi-definite, 
although it is both Hermitian and norm-preserving.
This divergence is shown in the inset of Fig.~\ref{fig:scaling}.
Upon including the stochastic terms, positivity is recovered on average, but
the numerical convergence of the Monte Carlo calculations is greatly retarded, 
particularly for long-time simulations.
Here we demonstrate that this problem can be largely mitigated by including a
classical portion of the real part of the influence functional in the
hierarchy.\cite{stockburger99}
Then the reduced density matrix is described by the standard
high-temperature hierarchy equations,
\begin{align}
   \frac{\partial}{\partial t} \hat \rho_n(\xi;t) 
      & = -\left(\frac{i}{\hbar} \tilde H(\xi; t)^\times + n \omega_c\right) \hat \rho_n(\xi; t)
      -\frac{i}{\hbar}\hat V^\times \hat \rho_{n+1}(\xi; t)
      - \frac{n\lambda}{\hbar}\left( \omega_c
                  \hat V^\circ  + i\frac{2}{\beta'}\hat V^\times
                  \right)\hat \rho_{n-1}(\xi; t)
   \label{eq:splotch}
\end{align}
and the weakened noise obeys the modified autocorrelation,
\begin{align}
   \langle \xi(t) \rangle & = 0 \nonumber \\
   \langle \xi(t) \xi(t') \rangle & = 
      \hbar \left[ K_r(t-t') - \gamma(t-t') \right]
      \;,\label{eq:splotch_noise}
\end{align}
where the classical friction function is defined by
\begin{align}
   \gamma(t) = \frac{2}{\pi\hbar\beta'} 
   \int_0^\infty d\omega\: \frac{J(\omega)}{\omega}
       \cos(\omega t)  
        = \frac{2 \lambda}{\hbar \beta'} e^{-\omega_c t}
   \;. 
\end{align}
In this approach, the temperature of the classical bath, $\beta'= 1/(k_b T')$ 
is a free parameter that can take on any value 
in the physically allowable range of $0 \le T' \le T$. 
Outside of this range the noise becomes complex, which is highly undesirable
from a numerical standpoint.
Setting $T'=0$, recovers the original Eq.~\ref{eq:hierarchy} 
proposed in Refs.~\onlinecite{Tanimura2006,zhou05} where the entire real part
of the bath correlation function is described by the noise.
In the opposite limit, when $T'=T$ then only the quantum corrections 
to the correlation function are treated stochastically.
At sufficiently high temperature where $K_r(t) = \gamma(t)$, this 
case reduces to the standard deterministic HEOM.
Eqns.~\ref{eq:splotch} and~\ref{eq:splotch_noise} constitute the central
formal developments of this work.

The impact of the choice of reference temperature on the convergence 
properties of the sHEOM is shown in more detail 
in Fig.~\ref{fig:scaling} for an unbiased two level system coupled 
to two independent, identical baths at $300$ K.
For $T'/T=1$, the temperature dependent term in the deterministic 
hierarchy (last term of Eq.~\ref{eq:splotch}) accounts for almost all of 
$K_r(t)$ so that convergence requires many levels of the hierarchy, 
but only few samples of the noise.
As the reference temperature is lowered, the temperature-dependent 
deterministic component of the hierarchy carries less weight as more of 
the bath correlation function is described by the stochastic sampling.
The numerical convergence steadily requires more Monte Carlo samples, 
but fewer hierarchy tiers.
This behavior continues until the temperature dependent, deterministic term 
becomes negligible compared with the dissipative term such that the 
required number hierarchy tiers is constant for $T'/T \lesssim 0.4$.
However, near $T'=0$ (Eq.~\ref{eq:hierarchy}) the noiseless hierarchy becomes
unstable as seen in the inset of Fig.~\ref{fig:scaling} and the required 
number of Monte Carlo samples increases rapidly.
We have found that the choice $\beta' = 2/\omega_c$,
(provided, of course, that $T'\le T$),
which equates the prefactors in the last term of Eq.~\ref{eq:splotch}, 
generally preserves the positivity of the reduced density matrix
and allows for quite stable simulations over long timescales, while 
also keeping the depth of the hierarchy to a minimum.
For example, this choice allows for
converged results with nearly two orders of magnitude fewer Monte Carlo 
samples than in Eq.~\ref{eq:hierarchy} without increasing the required 
number of hierarchy levels.
The quantitative behavior seen in Fig.~\ref{fig:scaling} is somewhat 
system dependent, but we have found a similar qualitative trend for 
all cases studied so far.
We will use the value $\beta=2/\omega_c$ in all of the computations 
presented below.

The numerical integration of the sHEOM 
requires only a minor modification to the standard numerical approach to 
the HEOM. 
First a realization of the stochastic noise trajectory is generated
that is consistent with the autocorrelation function given in 
Eq.~\ref{eq:splotch_noise}
The numerical generation of Gaussian noise with an arbitrary correlation
function is discussed in Appendix~\ref{app:noise}.
Following this, the stochastic Hamiltonian (Eq.~\ref{eq:stochastic_H})
is formed and the integration of the hierarchy equations proceeds as usual.
This procedure is then repeated until the dynamics of the 
reduced density matrix is converged to acceptable accuracy.
Typically this is reached within $10^4-10^6$ realizations of the noise.

\section{Numerical Results}
\label{sec:results}

\subsection{Two level systems}

\begin{figure}
   \includegraphics*[width=0.5\textwidth]{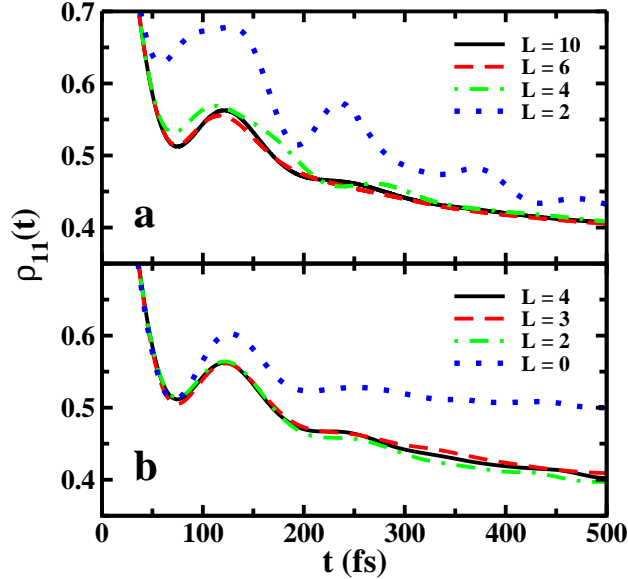}
   \caption{
      The convergence of the standard (a) and stochastic (b) 
      HEOM with respect to the number of tiers, $L$, 
      at a temperature of $T=300$ K.
      The bias, electronic coupling and system-bath coupling
      are $100$ cm$^{-1}$ while the cutoff frequency is $53$ cm$^{-1}$.
      The sHEOM results are computed with $N=10^4$ trajectories 
      at each tier.
   }
   \label{fig:TLS_300}
\end{figure}

In order to demonstrate the efficacy of the sHEOM approach, 
we first present benchmark studies of two level systems
for which results from the standard hierarchy can be independently 
computed.\cite{strumpfer12}
We consider the population dynamics in the biased two level system 
studied in Ref.~\onlinecite{ishizaki09b},
where each of the sites is coupled to an independent, identical bath.
The system Hamiltonian is given by 
$\hat H_0=\Delta \hat \sigma_z + J \hat \sigma_x$ 
where $\Delta= 50$ cm$^{-1}$ and $J = \lambda = 100$ cm$^{-1}$,
with the cutoff frequency $\omega_c = 53$ cm$^{-1}$.
Fig.~\ref{fig:TLS_300}a displays the standard hierarchy dynamics 
at $300$ K where the high temperature approximation is completely valid
and no Matsubara terms are required.
The sHEOM results in Fig.~\ref{fig:TLS_300}b are very well
behaved, and already at $L=0$ most of the short time dynamics are accurately
captured.
At this level, the dynamics are similar to the Haken-Strobl model 
except that the noise is colored and constructed so as to correctly describe
the real part of the bath correlation function. 
Since the dissipative effects of the bath are completely
neglected at $L=0$, the sites are equally populated at equilibrium.
Nevertheless, this approach has recently been used to study diffusion 
processes in large unbiased systems containing hundreds of 
sites,\cite{zhong11,chen13}
and is valid as long as the temperature is larger than the bandwidth of the
system.
In Fig.~\ref{fig:TLS_300}, it is readily seen that the $L=2$ results in 
the stochastic approach are almost completely converged while 
the corresponding standard HEOM results display highly erratic behavior.
At least $L=6$ is required for the standard hierarchy to obtain results of
comparable accuracy to the sHEOM approach at $L=3$. 
Indeed, small differences are seen in the standard hierarchy results up 
to $L=10$.
It should be noted that even for $L=10$, 
the standard HEOM is more efficient than the stochastic approach at any level,
but this advantage is quickly lost as the system size increases
or the temperature is lowered.
Additionally, one could set the temperature of the classical bath
to the physical temperature ($T'=T$ in Eq.~\ref{eq:splotch}) in which case 
the standard hierarchy is recovered since the low temperature corrections 
are nearly negligible for these parameters.

\begin{figure}
   \includegraphics*[width=0.5\textwidth]{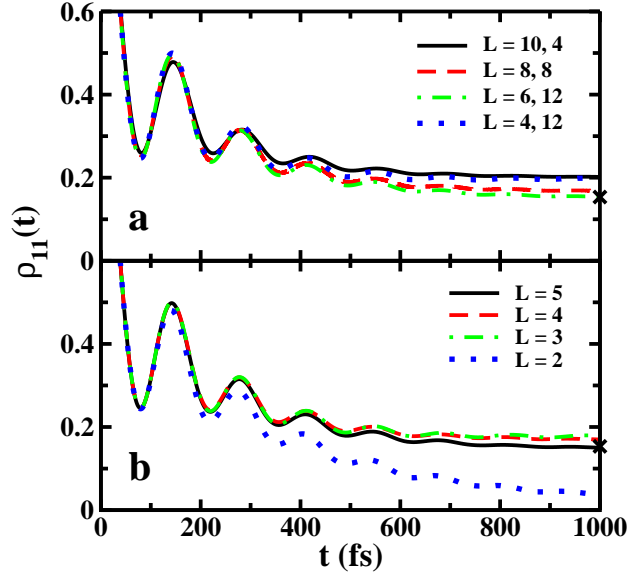}
   \caption{
      The convergence of the standard (a) and stochastic (b) 
      HEOM at a temperature of $T=10$ K with respect to the number
      of tiers and Matsubara terms (in (a)).
      The remaining parameters are the same as in Fig.~\ref{fig:TLS_300}.
      The stochastic calculations required $10^5$ Monte Carlo samples
      for convergence.
      The cross-symbols on the ordinate denote the exact equilibrium
      values.
   }
   \label{fig:TLS_10}
\end{figure}

A more interesting benchmark case is displayed in Fig.~\ref{fig:TLS_10} 
for the same system as in Fig.~\ref{fig:TLS_300} except with the temperature 
lowered from $300$ K to $10$ K. 
Fully converged results for the standard HEOM with respect to the 
number of Matsubara terms were possible for $4$ and $6$ hierarchy
tiers, but memory requirements limited convergence at level $L=8$ to 
$8$ Matsubara terms, and at $L=10$ to only $4$ Matsubara terms.
As can be seen by comparing the long-time behavior of the dynamics 
to the exact equilibrium value obtained from imaginary time path integral
calculations,\cite{moix12} most of these results are not fully converged.
The closest result to the exact limit for the standard hierarchy is 
$L=6$ with $12$ Matsubara terms,
although without being able to converge the $L=8$ calculations it is impossible
to know {\it a priori} if this is an acceptable result.
In contrast, the sHEOM approach is completely free of such demands.
Fully converged results are readily obtained as shown in 
Fig.~\ref{fig:TLS_10}b.
In addition as seen previously in Fig.~\ref{fig:TLS_300}, 
the sHEOM calculations are nearly converged at a lower tier
than in the standard approach.
In this case, the computational cost of the standard hierarchy with 
Matsubara terms at $L=6$ becomes comparable to that of the hybrid approach.
In the former case, one is required to integrate a single deterministic 
system of equations containing $\sim 10^6$ density matrices, 
while in the latter one must compute $10^5$ Monte Carlo samples, 
but the system of equations contains only $\sim 20$ matrices.
This scaling becomes even more favorable as the system size increases.

\subsection{Energy transfer rates}

As seen above, the sHEOM approach is capable of generating 
the numerically exact dynamics at both high and low temperatures,
as well as the correct equilibrium limit. 
Now we turn to simulations that are difficult, if not impossible, 
to carry out with the standard hierarchy, but become straightforward
when using the hybrid approach.
We first consider energy transfer in the two level donor-acceptor 
system considered in Ref.~\onlinecite{ishizaki09b}.
The bias, electronic coupling, and cutoff frequency
are $\Delta=50$ cm$^{-1}$, $J = 20$ cm$^{-1}$, and $\omega_c=53$ cm$^{-1}$,
respectively.
For this relatively weak value of the electronic coupling, the 
F{\"o}rster rates provide a reasonable approximation to the exact rates.
In Ref.~\onlinecite{ishizaki09b} the temperature was fixed at $300$ K
and the energy transfer rate from the higher lying electronic state 
to the lower state was computed as a function of the reorganization energy.
However, the reorganization energy is generally not an experimentally
tunable parameter.
Here, we fix the reorganization at the maximum energy transfer rate 
observed in Ref.~\onlinecite{ishizaki09b} ($\lambda=20$ cm$^{-1}$) 
and scan the temperature from $1$ to $1000$ K.
In every case, the results were converged with $L=2$ hierarchy tiers
and $10^4$ Monte Carlo samples.
The rates, $k$, are computed by fitting the long-time population 
dynamics to a kinetic model for the energy transfer dynamics between the
two sites.\cite{craig07}
In this model, the population of the initially excited state
is given by,
\begin{equation}
   P_1(t) = \frac{\chi_{\rm eq}+e^{-(1+\chi_{\rm eq})k t}}{1+\chi_{\rm eq}}
\end{equation}
where $\chi_{\rm eq} = P_1^{\rm eq}/P_2^{\rm eq}$ is the ratio of the 
equilibrium populations computed from imaginary time path integral
calculations.\cite{moix12}
In the results presented in Ref.~\onlinecite{ishizaki09b},
the accuracy of the F{\"o}rster rates improved for very large
or very small values of the reorganization energy.
Here, the F{\"o}rster rates are seen to systematically overestimate the exact
results for all temperatures. 
This error can be largely corrected by including the 
fourth order correction in the electronic coupling to the F{\"o}rster rates, 
as shown in previous studies of diffusion-limited electron
transfer.\cite{cao00,wu13}
\begin{figure}
   \includegraphics*[width=0.5\textwidth]{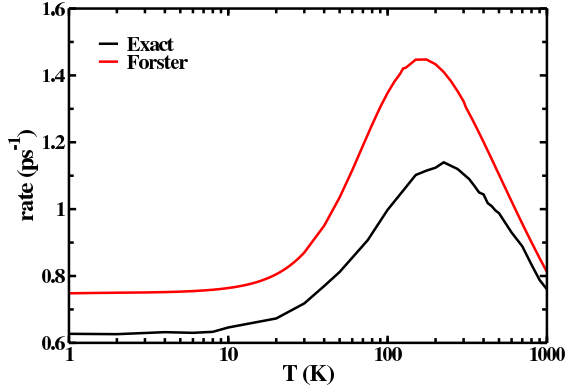}
   \caption{
      The energy transfer rates computed from the exact sHEOM 
      simulations (black) and the standard F{\"o}rster 
      rates (red).
      The bias is $\Delta=50$ cm$^{-1}$ and the electronic coupling,
      $J = 20$ cm$^{-1}$, with the cutoff frequency $\omega_c=53$ cm$^{-1}$
      and reorganization energy $\lambda=20$ cm$^{-1}$.
      The exact results are computed with $L=2$ levels of the hierarchy
      and $10^4$ Monte Carlo samples.
   }
   \label{fig:rates}
\end{figure}

\subsection{Entanglement dynamics}

\begin{figure}
   \includegraphics*[width=0.5\textwidth]{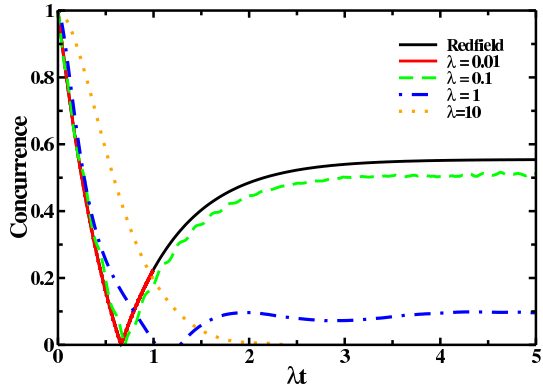}
   \caption{
      The exact concurrence between two qubits compared with the 
      results from the secular Redfield equation.
      The solid red, dashed green, dot-dashed blue, and dotted orange lines
      display the results for reorganization energies of 
      $\lambda/J = 0.01$, $0.1$, $1$, and $10$, respectively.
      The exact $\lambda=0.01J$ results are indistinguishable from 
      the Redfield results on this scale and for clarity are shown only 
      until $\lambda t=1$.
      $10^6$ Monte Carlo samples were used in each of the stochastic
      hierarchy calculations.
      In the scaled units of time, $\lambda t$, the Redfield dynamics are
      independent of $\lambda$.
   }\label{fig:concurrence}
\end{figure}

Finally, we present results on the entanglement dynamics of two
qubits at near zero temperature across a wide range of system-bath
coupling strengths. 
The system Hamiltonian is given by 
$\hat H_0 = \omega_0\left(\hat \sigma_1^z + \hat \sigma_2^z\right) 
+ J\hat \sigma_1^x \hat \sigma_2^x$
and the two qubits are coupled to identical, independent baths through their
respective $\hat \sigma^x$ operators.
The system frequency and bath cutoff are $\omega_0=1.5J$ and $\omega_c = 3 J$,
respectively, in units where $J=1$ sets the energy scale.
This model was considered in Ref.~\onlinecite{dijkstra10},
and, although not shown, we have reproduced the high temperature 
dynamics shown therein.
The initial state is chosen to be a completely entangled state,
$\rho(0) = I + \hat \sigma_1^x \hat\sigma_2^x+\hat\sigma_1^y\hat\sigma_2^y
-\hat \sigma_1^z\hat\sigma_2^z$,
where $I$ denotes the identity matrix.
Fig.~\ref{fig:concurrence} displays results for the concurrence 
with the low temperature, $\beta J=50$ converged with $10^6$ 
Monte Carlo samples.
At very weak coupling, $\lambda=0.01 J$ the results of the secular Redfield 
equation are in excellent agreement with the exact numerical results.
However, the equilibrium state generated by the Redfield dynamics is always
that given by the Boltzmann distribution computed with respect to $\hat H_0$,
regardless of the system-bath coupling strength.  
Additionally, in the scaled units of time, $\lambda t$, the secular Redfield 
dynamics are independent of $\lambda$.
Nevertheless, for weak system-bath coupling, $\lambda=0.1 J$, 
the Redfield dynamics still provide an accurate approximation to the exact 
results.
The entanglement displays an initial death followed by a subsequent 
reappearance and slow equilibration.\cite{yu09}
In contrast to the Redfield results, 
the exact dynamics correctly demonstrate that the 
environment-induced decoherence steadily destroys the entanglement 
between the qubits as the system-bath coupling strength increases. 
In fact, for $\lambda=10 J$ there is no reappearance of the concurrence after
the initial death.
In this case the system-bath coupling is sufficiently strong so 
as to destroy all of the equilibrium entanglement.

\section{Conclusions}

In summary, a hybrid stochastic hierarchy equation of motion (sHEOM) approach
has been proposed that substantially extends the parameter regimes accessible
to the hierarchy formalism. 
The method combines many of the strengths of the standard hierarchy with
those of recently proposed stochastic methods.\cite{Tanimura2006,stockburger02}
By performing a Hubbard-Stratonovich transformation only on the 
real part of the bath correlation function,
the problematic temperature dependent terms in the hierarchy are 
exchanged for a Monte Carlo average over real noise trajectories.
This extension eliminates the hierarchy tiers that are required to account 
for the low temperature corrections in the standard approach.
Consequently the numerical cost of the sHEOM simulations is 
nearly independent of the temperature.
Additionally, treating the imaginary part of the bath correlation
function by the hierarchy cures many of the normalization difficulties
associated with the purely stochastic methods.\cite{stockburger02}
The introduction of the reference temperature in Eq.~\ref{eq:splotch} allows 
one to achieve an optimal balance between the stochastic and deterministic
components of the evolution that substantially improves the convergence
properties of the algorithm.
Numerical results were presented across a broad range of parameters,
including both the high and low temperature limits, as 
well as the strong to weak system-bath coupling regimes.
The computations of the energy transfer rates 
and the low temperature entanglement dynamics in two qubits presented
in the previous section are difficult, if not impossible, to obtain
with most other methods.
However, these calculations are very straightforward with 
the sHEOM approach presented here.
If the noiseless hierarchy is well-behaved then the Monte Carlo procedure 
generally converges rapidly, but even in cases where positivity is not 
ensured convergence is generally achieved within $10^6$ samples. 
Additionally, the use of the hybrid approach allows for a much lower
truncation of the hierarchy tiers which, in principle, allows one
to simulate larger systems than is possible with the standard hierarchy.
Although not discussed here, we have easily performed 
calculations on the light harvesting system LH2 across a broad range
of temperatures.

The Hubbard-Stratonovich transformation of the real part of the bath 
correlation function appearing in the influence functional 
is valid for arbitrary spectral densities. 
It is only the hierarchy treatment of the remaining imaginary part 
that demands a Drude-Lorentz form.
In principle, it should be possible to employ other spectral densities 
such as that of the underdamped Brownian oscillator\cite{Tanaka2009} 
or by decomposing more complicated spectral densities into a sum of 
Drude-Lorentz terms.\cite{kreisbeck12}
However, the hierarchy of equations will be more complicated.
In these cases, it may be more advantageous to adapt the procedure 
outlined above to QUAPI,\cite{stockburger98}
which suffers many of the same problems as the standard HEOM in the 
low temperature regime.
The extension of the hybrid approach to the QUAPI formalism would allow for
numerically exact simulations of the non-Markovian dynamics of 
open quantum systems for arbitrary spectral densities across a broad 
range of the parameter space.

\section{Acknowledgments}

This work was supported by the NSF (Grant No. CHE-1112825) and 
DARPA (Grant No.~N99001-10-1-4063).
J.~Moix has been supported by the Center for Excitonics,
an Energy Frontier Research Center funded by the US Department of Energy,
Office of Science, Office of Basic Energy Sciences under Award 
No.~DE-SC0001088.
We thank Jian Ma and Arend Dijkstra for useful discussions, 
as well as the secular Redfield
code used to generate the results in Fig.~\ref{fig:concurrence}.

\appendix

\section{Noise generation}\label{app:noise}

Apart from the integration of the HEOM, which is achieved through 
standard Runge-Kutta methods, the only additional computational requirement 
of the sHEOM approach is the generation of the stochastic process, $\xi(t)$.
There are numerous methods to generate Gaussian colored noise.
In this appendix, two approaches are briefly described that we have found
to be particularly useful.
The most straightforward approach originates from signal processing techniques 
and is based on filtering white noise through an appropriate kernel,
\begin{equation}
   \xi(t) = \int_0^t dt' k(t-t')\zeta(t') \;.
\end{equation}
The noise, $\zeta(t)$, is a standard Wiener process with zero mean, 
$\langle \zeta(t) \rangle = 0$ 
and autocorrelation, $\langle \zeta(t)\zeta(t')\rangle = \delta(t-t')$.
The filtering kernel, $k(t)$, is determined from the factorization
of the autocorrelation function,
\begin{equation}
   K_r(t-t') = \int dt'' k(t-t'')k(t''-t') \;.
\end{equation}
In practice, $k(t)$ is most efficiently constructed through a Cholesky
decomposition of the discretized kernel matrix $\bar K_{ij} = K_r(t_i-t_j)$
with $\bar k_{ij}$ defined accordingly, such that $\bar K = \bar k^T \bar k$.
A discretized sample of the desired noise sequence is then simply generated 
from the matrix-vector product $\vec \xi = \bar k^T \vec \zeta$.

While straightforward, the Cholesky approach encounters difficulties in
simulations where many time steps are needed so that 
the kernel matrix becomes large.
In this case, an alternative approach to generate the noise can 
be used that relies on a discretization of the correlation function 
in terms of its independent frequency components.\cite{zhong11,billah90}
Samples of the noise can be generated from the Fourier sum
\begin{equation}
   \xi(t) = \sqrt{\frac 2\pi}\sum_{n=1}^N 
            \left[J(\omega_n)\coth\left(\frac{\beta \omega_n}{2}\right) 
               \Delta \omega \right]^{1/2} \cos(\omega_n t + \phi_n)
               \label{eq:noise}
\end{equation}
where $\Delta \omega = \omega_{\rm max}/N$ and $\omega_n = n\Delta \omega$ 
with $\omega_{\rm max}$ chosen to be sufficiently large such that the spectral 
density has decayed to zero, $J(\omega_{\rm max}) = 0$.
The phases, $\phi_n$, are uniform random numbers generated on the interval
$(0,2\pi)$.
The noise sequence generated in this manner is periodic with a period of 
$2\pi/\Delta \omega$ which, obviously, must be at least twice the simulation 
time.
The summation in Eq.~\ref{eq:noise} is most efficiently performed through the 
use of a fast Fourier transform.

\end{document}